\documentclass[12pt]{article}
\usepackage[utf8]{inputenc}
\usepackage{authblk}
\usepackage{setspace}
\usepackage[margin=1.25in]{geometry}
\usepackage{graphicx}
\usepackage{subcaption}
\usepackage{amsmath}
\usepackage{lineno}
\usepackage{lipsum}

\usepackage[style=nejm, 
citestyle=numeric-comp,
sorting=none]{biblatex}
\begin{document}
\title{Elliptic flow of light nuclei in Au+Au collisions at $\sqrt{s_{NN}}$ = 14.6, 19.6, 27, and 54.4 GeV using the STAR detector}
\author[1,*]{Rishabh Sharma (for the STAR Collaboration)}
\affil[1]{Indian Institute of Science Education and Research, Tirupati 517507, India}
\affil[*]{rishabhsharma@students.iisertirupati.ac.in}

\onehalfspacing
\maketitle

\date{}

%%%%%% Abstract %%%%%%
\begin{abstract}

Loosely bound light nuclei are produced in abundance in heavy-ion collisions. There are two main possible models to explain their production mechanism - the thermal model and the coalescence model. The thermal model suggests that the light nuclei are produced from a thermal source, where they are in equilibrium with other species present in the fireball. However, due to the small binding energies, the produced nuclei are not likely to survive the high-temperature conditions of the fireball. The coalescence model tries to explain the production of light nuclei by assuming that they are formed at later stages by the coalescence of protons and neutrons near the kinetic freeze-out surface. The final-state coalescence of nucleons will lead to the mass number scaling of the elliptic flow ($v_2$) of light nuclei. This scaling states that the $v_2$ of light nuclei scaled by their respective mass numbers will follow very closely the $v_2$ of nucleons. Therefore, studying the $v_2$ of light nuclei and comparing it with the $v_2$ of protons will help us in understanding their production mechanism.

In this talk, we will present the transverse momentum ($p_{T}$) and centrality dependence of $v_2$ of $d$, $t$, and $^3\text{He}$ in Au+Au collisions at $\sqrt{s_{NN}}$ = 14.6, 19.6, 27, and 54.4 GeV. Mass number scaling of $v_2(p_T)$ of light (anti-)nuclei will be shown and physics implications will be discussed.

\end{abstract}
\newpage

\section{Introduction}
The study of light nuclei and their interaction in high-energy heavy-ion collisions has been a subject of active theoretical  and experimental investigations \cite{OLIINYCHENKO2021121754}. The production mechanism of light nuclei in heavy-ion collisions is not very well understood. There are two main models that describe this mechanism: the thermal model and the coalescence model. The thermal model suggests that light nuclei are formed near the chemical freezeout (CFO) surface along with other hadrons \cite{Andronic_Braun-Munzinger_Redlich_Stachel_2018}. However, the low binding energies of light nuclei make it unlikely that they will be able to sustain the high temperature at CFO. The coalescence model, on the other hand, suggests that light nuclei might be formed by the coalescence of nucleons at the later stages of evolution of the system \cite{PhysRev.129.836}. This will result into the mass number scaling whereby elliptic flow of light nuclei scaled by their respective mass numbers follows closely to elliptic flow of protons \cite{YAN200650}. Therefore, by examining the collective flow of light nuclei, valuable insights can be gained into how they are produced in heavy-ion collisions. \\
In the following sections, we will report the elliptic flow ($v_2$) of d, t, and $^3$He in Au+Au collisions at $\sqrt{s_{NN}}$ = 14.6, 19.6, 27, and 54.4 GeV. We will also report the results of the centrality dependence study of $v_2$ of d in $\sqrt{s_{NN}}$ = 19.6, 27, and 54.4 GeV. Finally, we will show the results from the mass number scaling study of $v_2(p_T)$ of light nuclei.

\section{Analysis details}
The data presented in these proceedings was collected during the second phase of the Beam Energy Scan (BES-II) program, conducted by the STAR experiment at RHIC. Light nuclei identification was done using the Time Projection Chamber (TPC) \cite{ACKERMANN2003624} and the Time of Flight (TOF) \cite{LLOPE2005306} detectors. TPC serves as the main tracking detector in the STAR experiment and relies on the measurement of specific ionization energy loss (dE/dx) within a large gas volume to identify and track various charged particles. The TOF detector, on the other hand, enables the identification of particles of interest by imposing a constraint on their mass-square ($m^2$).\\
Elliptic flow, $v_2$, is the second order Fourier coefficient of the azimuthal distribution of the produced nuclei relative to the reaction plane of the Au+Au collision. Since it is not feasible to directly measure the reaction plane angle in an experimental setup, we employ the TPC to construct the second order event plane angle ($\Psi_{2}$) as a substitute for the reaction plane angle \cite{PhysRevC.58.1671}. In the next section, we will discuss the results of $v_2$ of $d$, $t$, and $^3\text{He}$.

\section{Results}
\subsection{Elliptic flow of light nuclei}
\begin{figure}[!ht]
     \centering
     \begin{subfigure}[b]{0.24\textwidth}
         \centering
         \includegraphics[width=\textwidth]{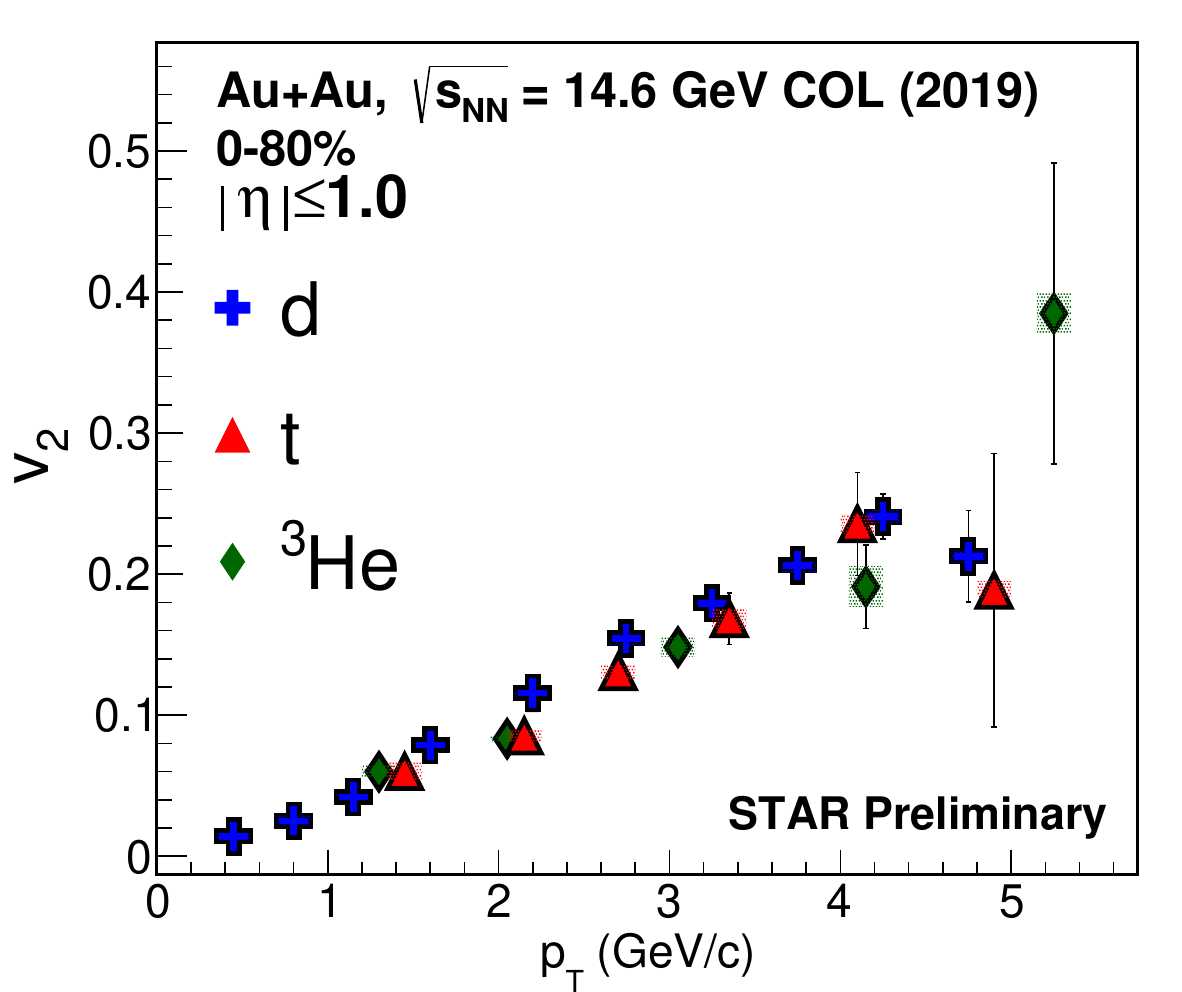}
     \end{subfigure}
     \hfill
     \begin{subfigure}[b]{0.24\textwidth}
         \centering
         \includegraphics[width=\textwidth]{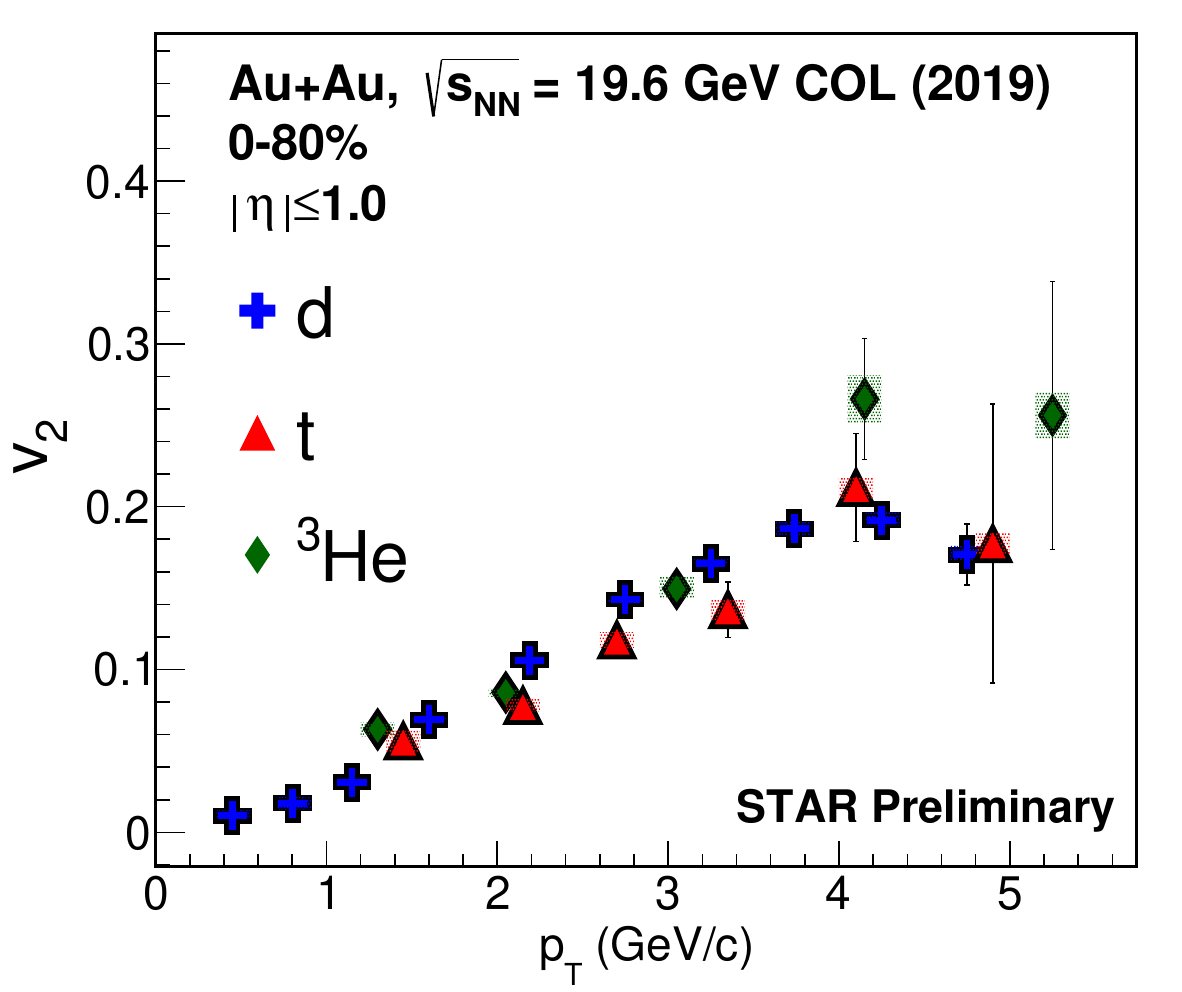}
     \end{subfigure}
     \hfill
     \begin{subfigure}[b]{0.24\textwidth}
         \centering
         \includegraphics[width=\textwidth]{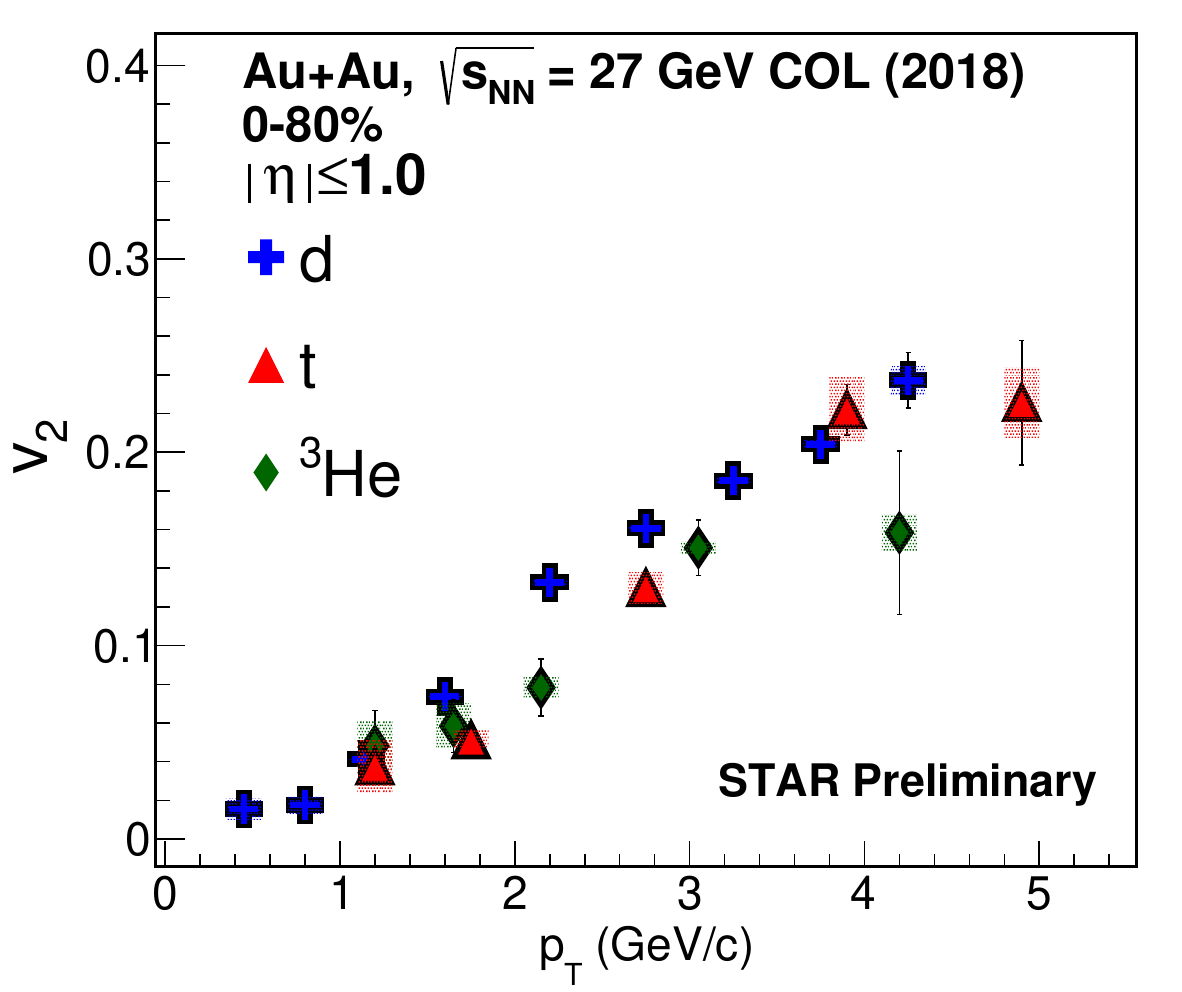}
     \end{subfigure}
     \begin{subfigure}[b]{0.24\textwidth}
         \centering
         \includegraphics[width=\textwidth]{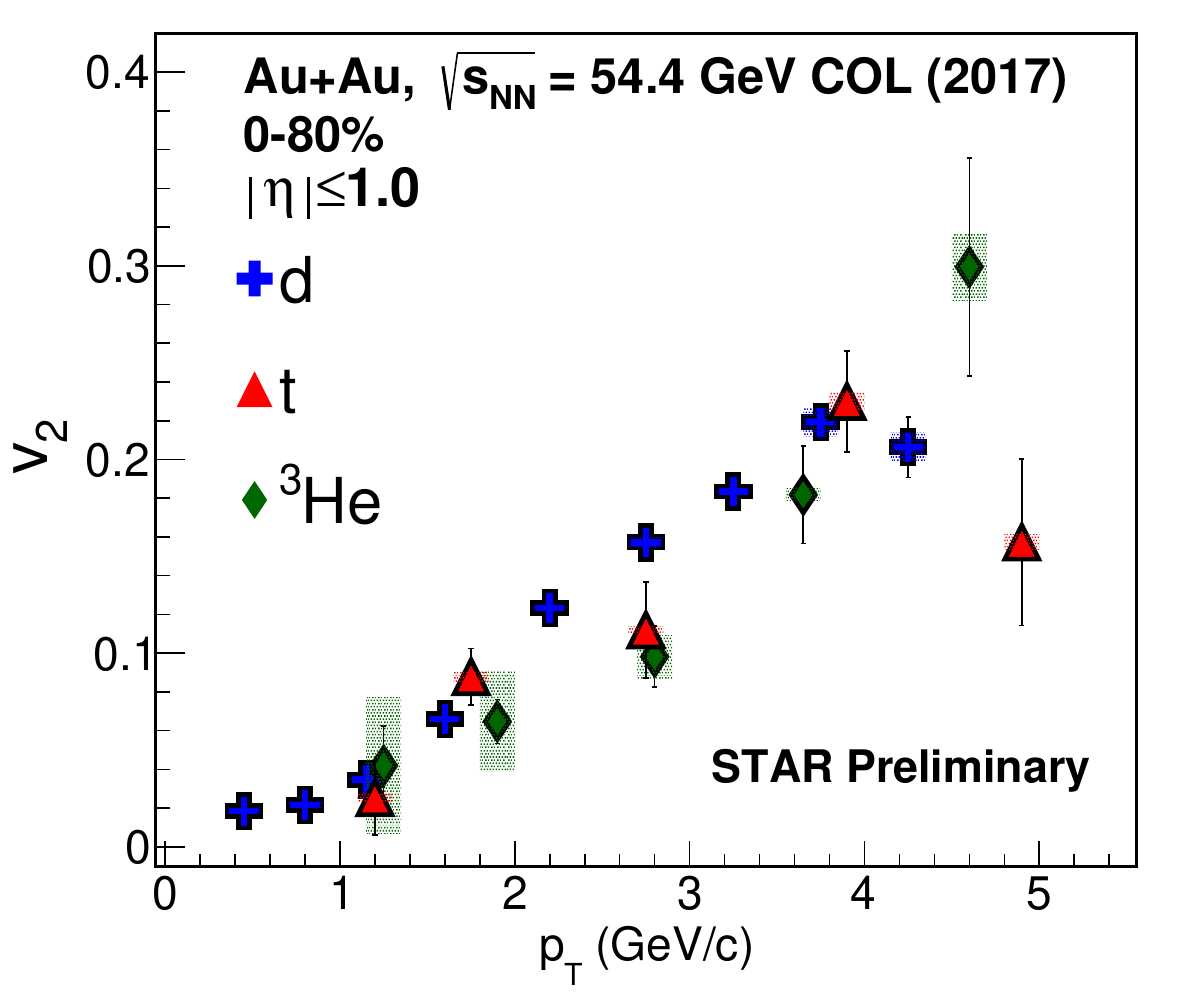}
     \end{subfigure}
        \caption{$v_2(p_T)$ of light nuclei ($d$, $t$, and $^3\text{He}$) in 0-80\% centrality Au+Au collisions at $\sqrt{s_{NN}}$ = 14.6, 19.6, 27, and 54.4 GeV. Vertical lines and shaded bands at each marker represent statistical and systematic uncertainties, respectively.}
        \label{fig:flow}
\end{figure}

Figure \ref{fig:flow} shows $v_2$ as a function of $p_T$ in 0-80\% centrality Au+Au collisions at $\sqrt{s_{NN}}$ = 14.6, 19.6, 27, and 54.4 GeV. A monotonous increase with $p_T$ in $v_2$ of light nuclei is observed across all four center-of-mass energies. 

\subsection{Centrality dependence of $v_2$}
\begin{figure}[!ht]
     \centering
     \begin{subfigure}[b]{0.30\textwidth}
         \centering
         \includegraphics[width=\textwidth]{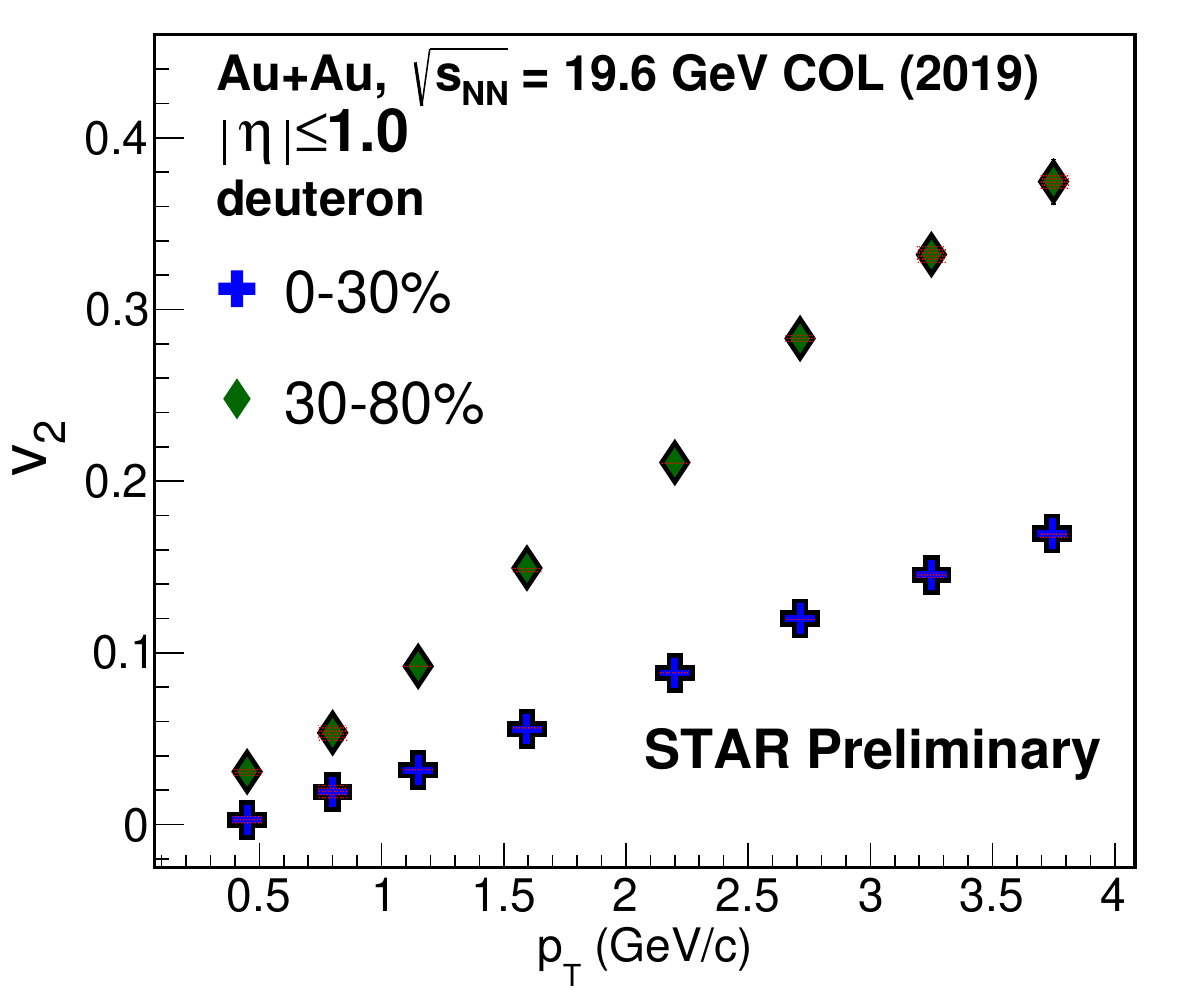}
     \end{subfigure}
     \hfill
     \begin{subfigure}[b]{0.30\textwidth}
         \centering
         \includegraphics[width=\textwidth]{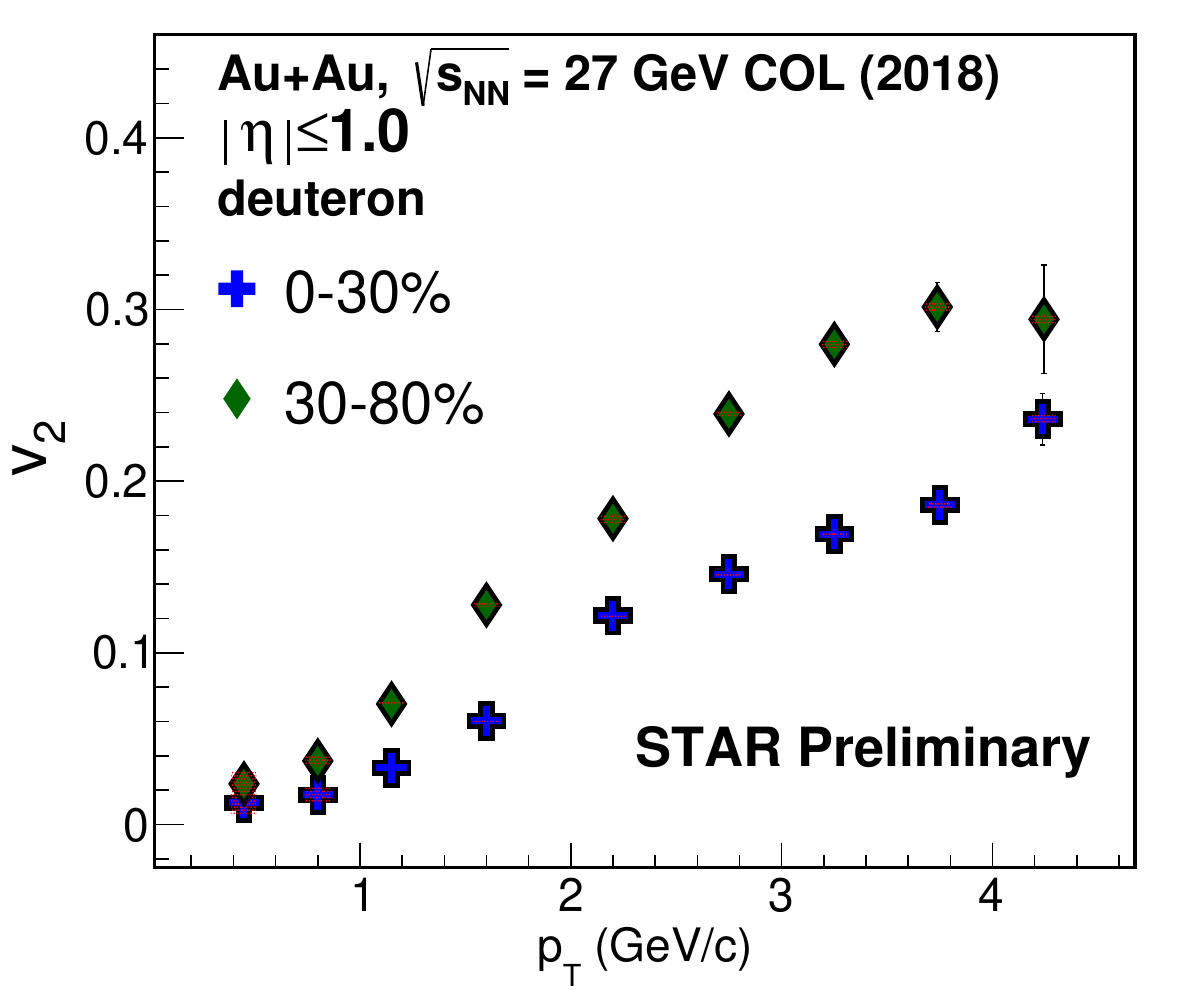}
     \end{subfigure}
     \hfill
     \begin{subfigure}[b]{0.30\textwidth}
         \centering
         \includegraphics[width=\textwidth]{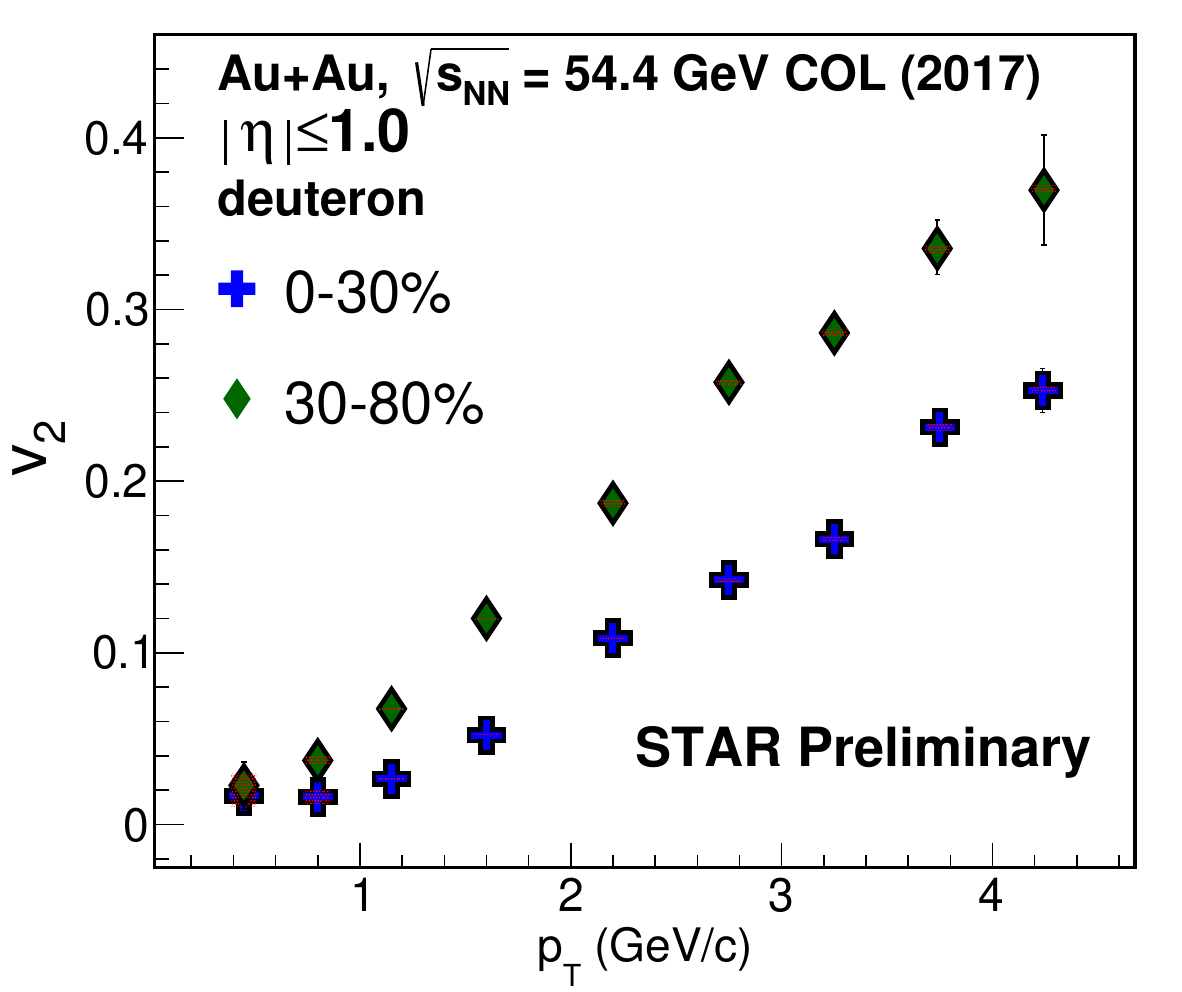}
     \end{subfigure}
        \caption{Centrality dependence of $v_2$ of $d$ as a function of $p_T$ in Au+Au collisions at $\sqrt{s_{NN}}$ = 19.6, 27, and 54.4 GeV. Vertical lines and shaded bands at each marker represent statistical and systematic uncertainties, respectively.}
        \label{fig:cent}
\end{figure}

Centrality dependence of $v_2$ of $d$ is shown in Fig. \ref{fig:cent}. The nuclei $v_2$ is measured in two centrality ranges 0-30\% and 30-80\% for Au+Au collisions at $\sqrt{s_{NN}}$ = 19.6, 27, and 54.4 GeV. It is noted that peripheral collisions exhibit higher $v_2$ values compared to more central collisions. This observation can be attributed to the greater spatial anisotropy in peripheral collisions as opposed to central collisions.

\subsection{Mass number scaling}
According to the coalescence model, assuming that protons and neutrons behave in the same way, for a light nuclei $N$ with mass number $A$, we expect $v_{2,N}(p_T) \approx Av_{2,p}(p_T/A)$, where $v_{2,p}$ is elliptic flow of protons  \cite{PhysRevC.94.034908,YAN200650,PhysRevC.76.054910}. The phenomenon is referred to as mass number scaling. Figure 3 shows the comparison of $v_2/A$ of light nuclei as a function of $p_T/A$ (where $A$ is the mass number of the nuclei) with $v_2/A$ of proton (where $A=1$). Proton $v_2$ has been fitted with a third-order polynomial. The bottom panel in each plot shows the ratio between the $v_2/A$ of light nuclei and the fit to proton $v_2$. It is observed that $v_2$ of light nuclei deviates from mass number scaling by 20-30\%. However, additional model studies are required to conclude whether the coalescence model is the dominant production mechanism of light nuclei. 

\begin{figure}[!ht]
     \centering
     \begin{subfigure}[b]{0.24\textwidth}
         \centering
         \includegraphics[width=\textwidth]{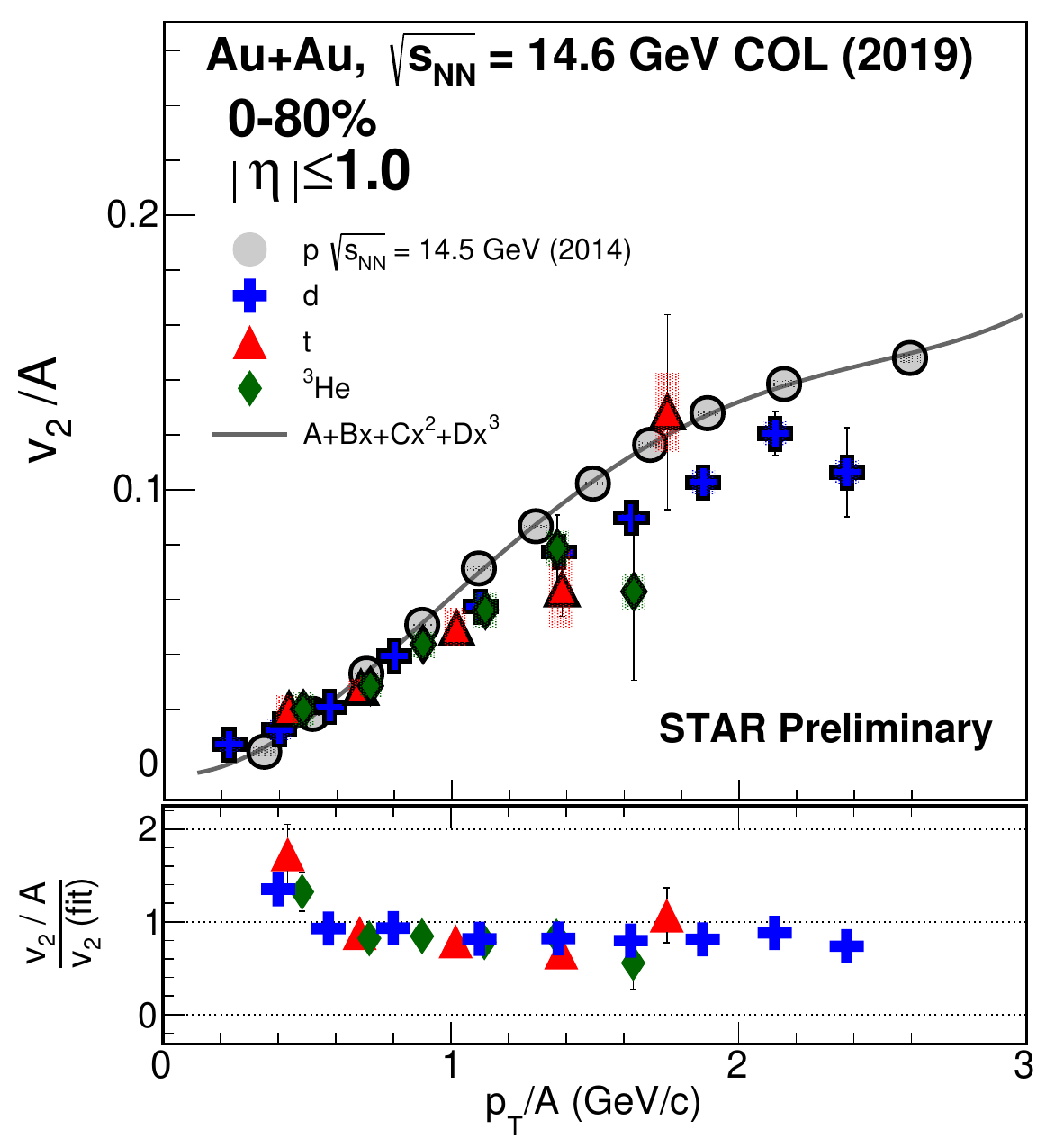}
     \end{subfigure}
     \hfill
     \begin{subfigure}[b]{0.24\textwidth}
         \centering
         \includegraphics[width=\textwidth]{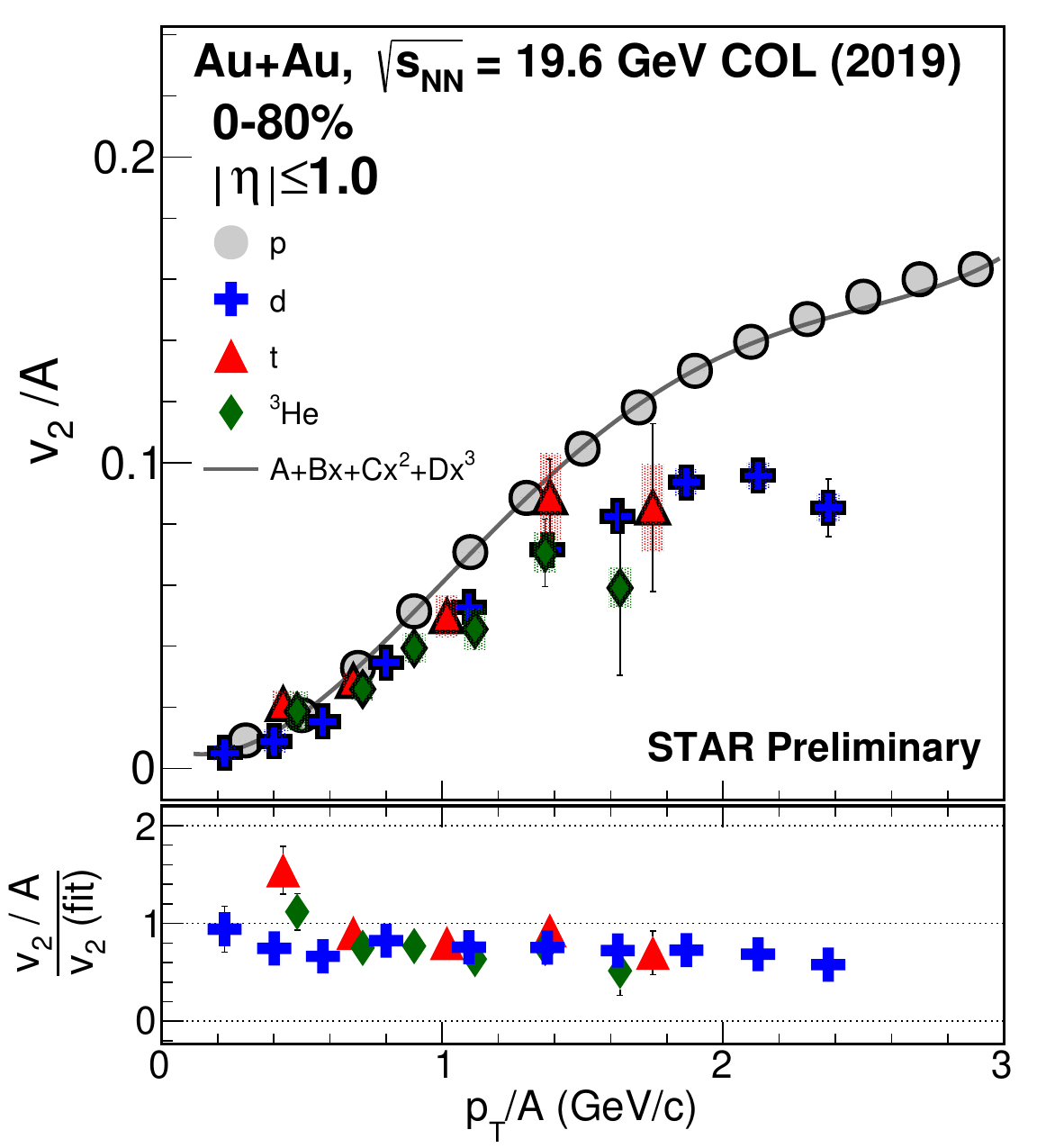}
     \end{subfigure}
     \hfill
     \begin{subfigure}[b]{0.24\textwidth}
         \centering
         \includegraphics[width=\textwidth]{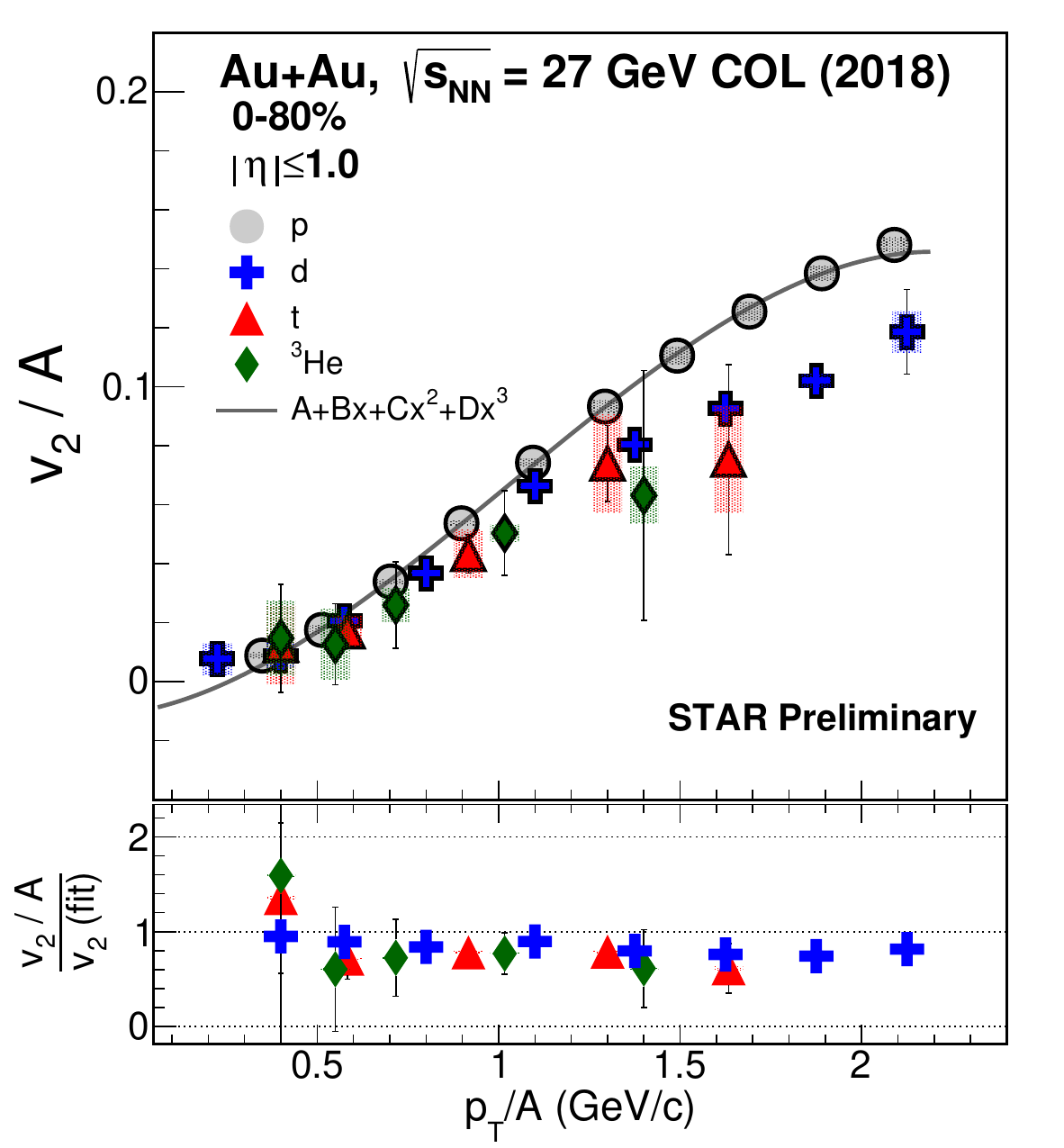}
     \end{subfigure}
     \begin{subfigure}[b]{0.24\textwidth}
         \centering
         \includegraphics[width=\textwidth]{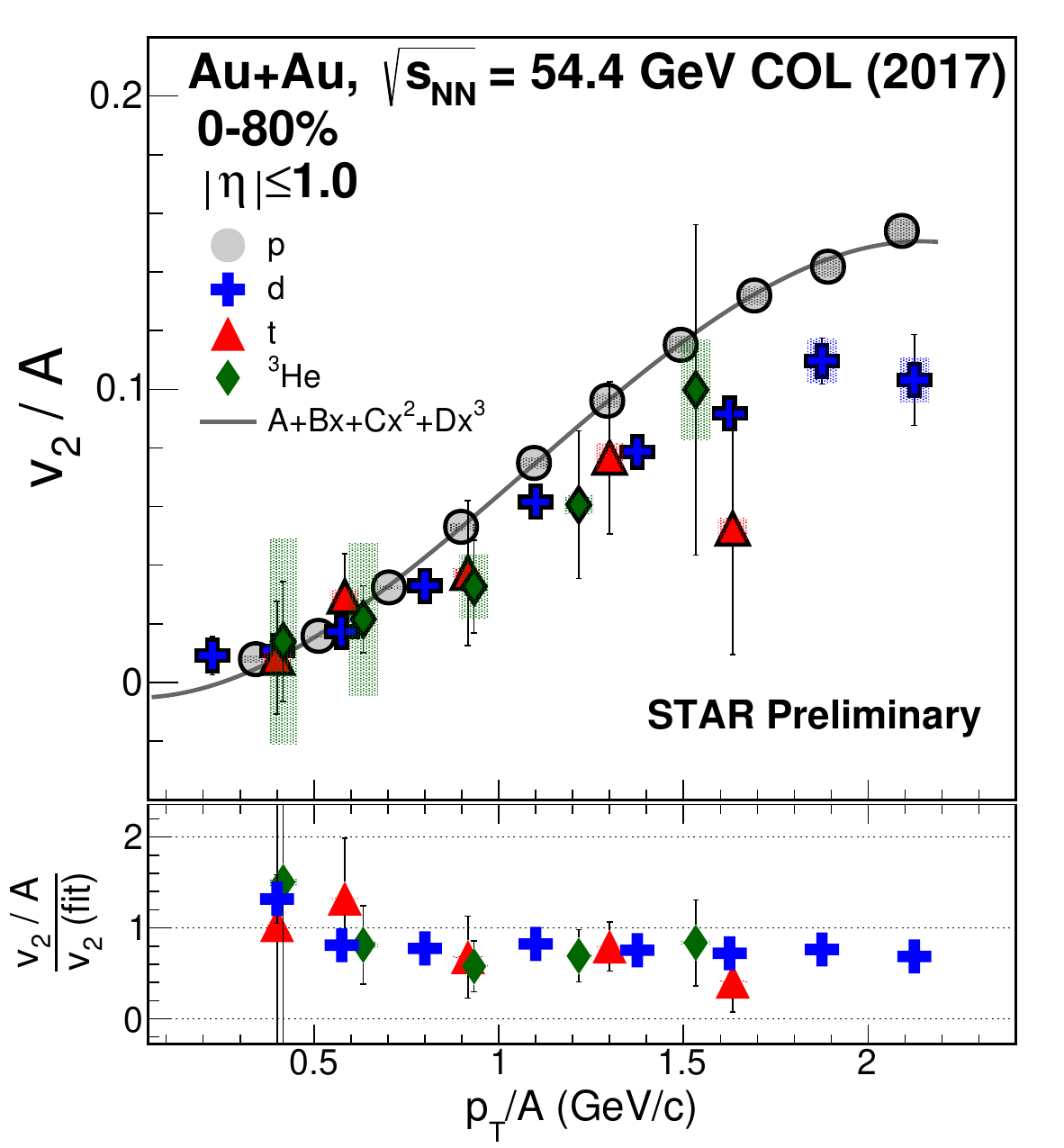}
     \end{subfigure}
        \caption{Mass number scaling of $v_2/A$ of light nuclei as a function of $p_T/A$ in minimum bias Au+Au collisions at $\sqrt{s_{NN}}$ = 14.6, 19.6, 27, and 54.4 GeV. Vertical lines and shaded bands at each marker represent statistical and systematic uncertainties, respectively.}
        \label{fig:mns}
\end{figure}

\section{Conclusion}
In summary, we have reported the $v_2$ of $d$, $t$, and $^3\text{He}$ in Au+Au collisions at $\sqrt{s_{NN}}$ = 14.6, 19.6, 27, and 54.4 GeV. A monotonic rise of light nuclei $v_2$ with $p_T$ is observed for all light nuclei species and studied energies. $v_2$ of $d$ is observed to show a strong centrality dependence being higher for peripheral collisions compared to central collisions. This behaviour can be attributed to the fact that peripheral collisions have higher spatial anisotropy compared to the central ones. In addition, it is also observed that $v_2$ of light nuclei deviates from mass number scaling by 20-30\%.

\printbibliography

\end{document}